\documentclass[%
 reprint,
 amsmath,amssymb,
 aps,
]{revtex4-2}

\usepackage{graphicx}
\usepackage{dcolumn}
\usepackage{bm}

\usepackage{hyperref}

\begin{document}

\preprint{APS/123-QED}

\title{Unveiling unique ultrafast nonlinearities in liquid-phase high-order harmonic generation}

\author{Wanchen Tao$^{1}$}
\author{Zhuang-Wei Ding\footnotemark$^{2,3}$}
\author{Lixin He$^{1}$}\email{helx\_hust@hust.edu.cn}
\author{Changlong Xia$^{4}$}
\author{Xingdong Guan$^{1}$}
\author{Xue-Bin Bian$^{2}$}\email{xuebin.bian@wipm.ac.cn}
\author{Pengfei Lan$^{1}$}\email{pengfeilan@hust.edu.cn}
\author{Peixiang Lu$^{1}$}\email{lupeixiang@hust.edu.cn}

\affiliation{%
        $^1$ Wuhan National Laboratory for Optoelectronics and School of Optical and Electronic Information, Huazhong University of Science and Technology, Wuhan 430074, China\\
        $^2$ Wuhan Institute of Physics and Mathematics, Innovation Academy for Precision Measurement Science and Technology, Chinese Academy of Sciences, Wuhan 430071, China\\
        $^3$ School of Physical Sciences, University of Chinese Academy of Sciences, Beijing 100049, China\\
        $^4$ College of Physics and Information Engineering, Shanxi Normal University, Taiyuan 030031, China\\}

\date{\today}

\begin{abstract}
High-order harmonic generation (HHG) provides a powerful optical tool for probing ultrafast dynamics on the attosecond timescale. While its mechanisms in gases and solids are well-established, understanding nonlinear optical responses in liquids remains challenging. The absence of long-range order in liquids questions the applicability of the existing HHG models developed in other media. Through combined experimental and theoretical investigations, we identify unique characters of liquid-phase HHG---spectral redshift and broadening, which are fundamentally distinct from both the gaseous and solid-state counterparts. Quantitative measurements and simulations of HHG in liquids illustrate a near linear dependence of harmonic redshift and broadening on the laser intensity, with the nonlinear response of water exceeding that of ethanol. The simulations reveal that these features arise from delocalized electronic states with energy loss in multiple scatterings and transient Stark shift during their transitions in laser fields. Meanwhile, we find that liquid polarity or hydrogen bond exerts decisive control over the transition dipole momentum distributions of delocalized states. Our findings establish a nonlinear spectral method for probing the internal network in liquids, paving the way for studying its role in chemical and biological processes.
\end{abstract}

\maketitle


\section{\label{sec:level1}Introduction}

High-order harmonic generation (HHG) has been widely explored in gaseous media \cite{mcpherson1987studies,Huillier1,Paul1,Hentschel1} and crystalline solids \cite{Ghimire1,Luu1,Ndabashimiye1} with well-established mechanisms such as the three-step model \cite{Corkum1,SFA} and Bloch electron dynamics \cite{Higuchi1,Vampa3,wu2015high,li2019reciprocal,li2023high}. These frameworks have not only established HHG as a sensitive spectroscopic tool but also deepened our understanding of light-matter interaction across different phases. In contrast, the study of HHG in liquids \cite{DiMauro1,Luu2,Svoboda1,Mondal1,mondal2,Ferchaud1,alexander1,xu1,yang1,Zeng1}—even though liquids are the most prevalent and functionally important phase of matter at the molecular scale—remains in its early stages. The disordered yet highly correlated nature of liquids, characterized by short-range molecular order and long-range structural disorder \cite{March1}, presents both a fundamental challenge and a unique opportunity. Liquids host electronic states that are neither fully localized, as in gases, nor perfectly delocalized, as in periodic solids. This intermediate regime renders liquids a compelling platform for probing strong-field and attosecond electron dynamics in environments governed by structural complexity.
	
	In liquids, delocalized electronic states underpin many essential liquid-phase phenomena, from solvation and proton transport to dielectric response and reactivity \cite{Siefermann1,Gei1,Sellberg1}. These delocalized states in liquids span fluctuating molecular networks, giving rise to nonlocal electron motions that are qualitatively different from those in either gas or crystalline systems. Understanding the real-time behavior of these states is essential for uncovering the microscopic origins of macroscopic liquid properties and for advancing fields ranging from femtochemistry to biophysics. However, directly accessing the behavior of these states under strong optical fields has remained a significant experimental and theoretical challenge.
	High-harmonic spectroscopy offers a powerful all-optical probe with inherent attosecond temporal resolution, and it has been successfully applied to reveal electron dynamics in both atomic gases \cite{itatani2004tomographic,chargemigration1,smirnova2009high,NC} and crystalline materials \cite{Vampa1,lu2019multielectron,li2020determination}. Yet in liquids, the spectral features of HHG—which are known to encode information about electronic structure and scattering—remain largely unexplored. Key questions remain open: How do delocalized electrons in liquids respond to intense optical fields? What roles do structural disorder and dynamic scattering play in shaping the emitted harmonics? And how can HHG be harnessed to probe these elusive electronic features?
	
	In this study, we investigate high-harmonic generation from bulk liquid samples driven by intense femtosecond laser fields. Through comparative measurements in water, ethanol, and gaseous argon, we identify universal intensity-dependent spectral redshifts and broadening in the liquid phase. These spectral features are consistently observed across different liquids but absent in the gas, indicating a phenomenon intrinsic to the liquid phase. To interpret these observations, we develop an \textit{ab initio} simulation framework to model laser-driven electron dynamics in disordered liquids. The spectral modulations in liquid HHG are attributed to scattering-dominated nonlocal electronic trajectories modulated by liquid-specific structural disorder. The interplay between delocalized-state electron motion in liquid and dynamic scattering emerges as the governing mechanism, contrasting sharply with Bloch oscillations in solids or recollision physics in gases. Particularly, the intensity-dependent spectral broadening in liquids is associated with the transient Stark shift of delocalized states. This observation underscores the critical role of liquid polarity or hydrogen bond network in the reshaping of electronic state distributions. Our findings establish a direct link between strong-field harmonic emission and the underlying electronic environment of liquids, demonstrating HHG as a powerful probe for accessing delocalized-state electron dynamics in complex, disordered media. This approach paves the way for attosecond spectroscopy in disordered and soft condensed phases, with potential implications for understanding charge transport, solvation, and light-driven chemical processes in complex environments.

\section{Spectral redshift and broadening in liquid HHG}
\begin{figure}[b]
\includegraphics[width=0.5\textwidth]{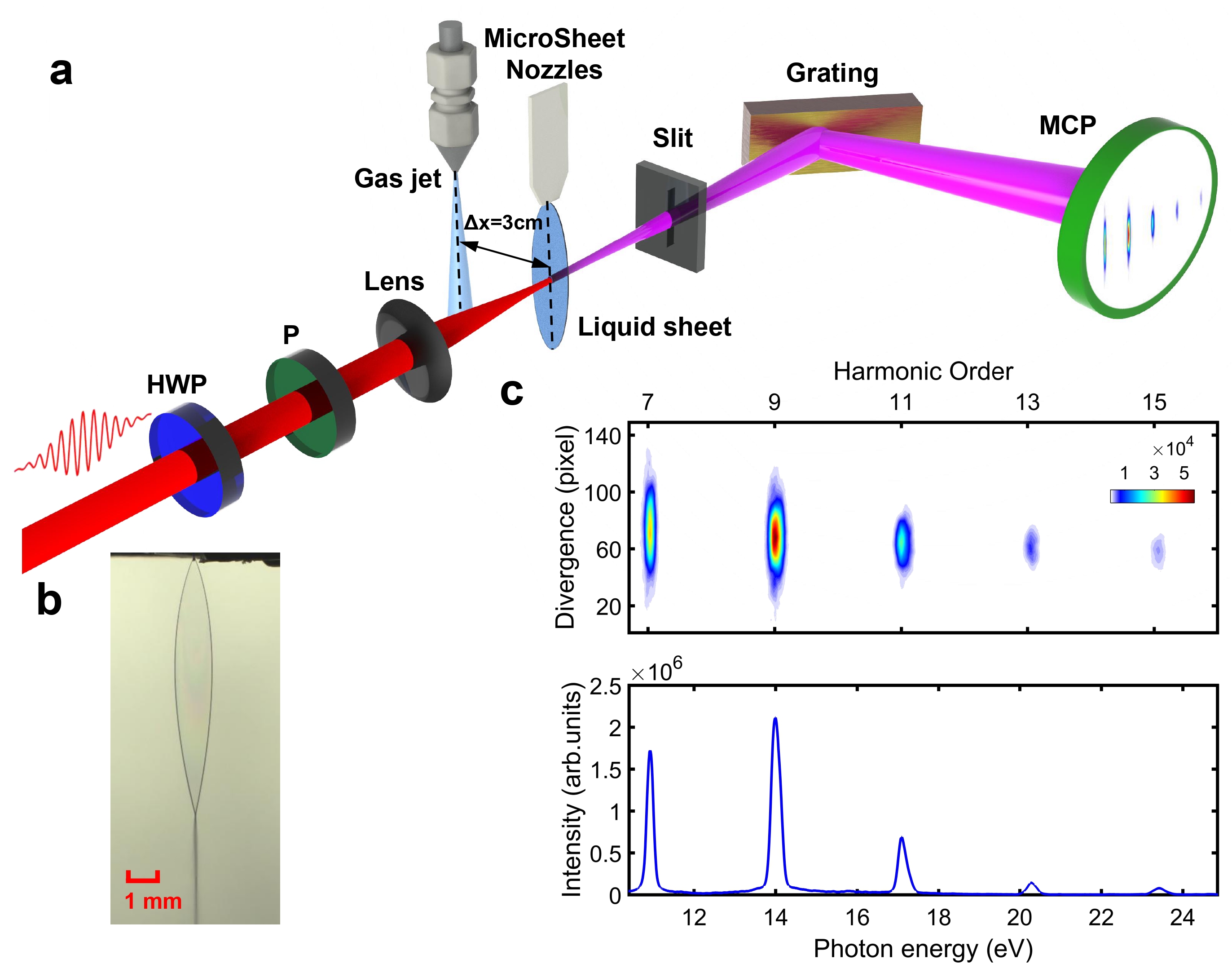}
\caption{\label{fig:1} \textbf{High-order harmonic generation from liquids.} \textbf{a,} Experimental setup. HWP: half-wave-plate, P: polarizer. MCP: microchannel plate. \textbf{b,} Photograph of the liquid flat-sheet. \textbf{c,} Typical spatio-spectrally resolved (upper) and spatially integrated (lower) harmonic spectra from the liquid water.}
\end{figure}

The experimental setup is illustrated schematically in Fig. \ref{fig:1}(a). A femtosecond laser pulse (50 fs duration, centered at 800 nm) is focused onto a liquid flat-sheet inside a vacuum chamber to generate high-order harmonics. Figure \ref{fig:1}(b) displays the image of the liquid flat-sheet produced by MicroSheet nozzles in our experiment. The generated harmonics were recorded by a homemade flat-field soft x-ray spectrometer (see Methods). Figure \ref{fig:1}(c) shows a typical spatio–spectrally resolved high-harmonic spectrum (upper) generated from liquid water, along with the corresponding spatially integrated harmonic signal (lower). The spectrum exhibits a series of odd harmonics extending up to the 15th order (H15), with wavelengths reaching below 60 nm. A distinct plateau is observed from H7 and H9, followed by a steep intensity drop in the cutoff region from H11 to H15.

Using this setup, we have systematically measured HHG spectra from liquid water and ethanol under varying laser intensities. In the experiment, the laser intensity was precisely controlled by rotating a half-wave plate relative to a polarizer. For comparison, HHG from argon gas was also measured by replacing the microfluidic chip with a gas supersonic nozzle.

\begin{figure*}
\includegraphics[width=0.75\textwidth]{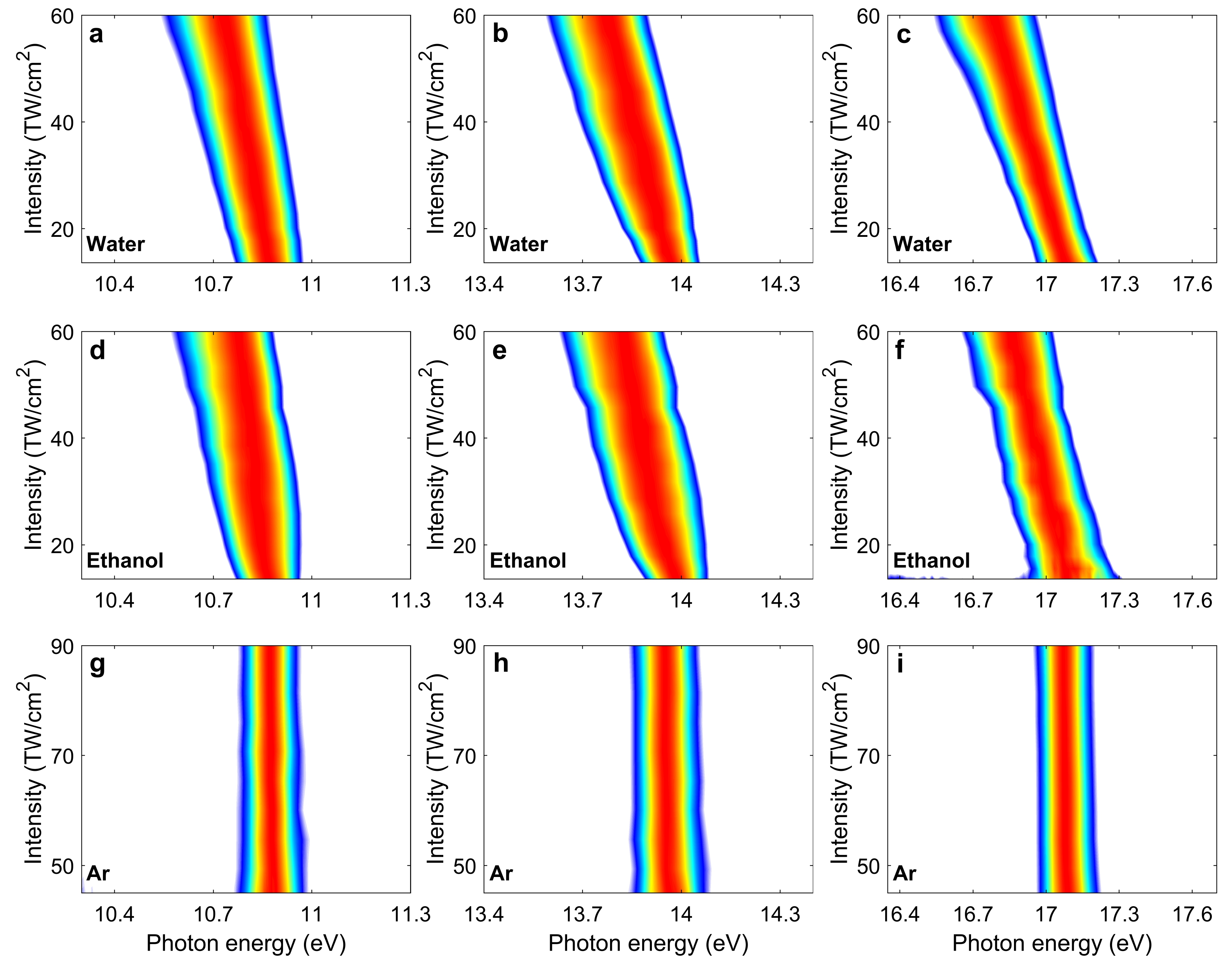}
\caption{\label{fig:2} \textbf{Laser-intensity-dependent harmonic spectra from liquids and gas.} \textbf{a-c,} Laser-intensity-dependent spectra of H7-H11 from liquid water. \textbf{d-f} and \textbf{g-h}, Same as \textbf{a-c}, but for liquid ethanol and argon gas, respectively.}
\end{figure*}

Figure \ref{fig:2} displays the harmonic spectra of H7-H11 from liquid water (top row), ethanol (middle row), and argon gas (bottom row) as a function of the laser intensity. Note that to avoid the plasma effects \cite{Ferchaud1}, we have restricted the laser intensity below 60~$\text{TW}\text{/cm}^{2}$ in the liquid HHG measurement. Moreover, at each intensity, the measured spectra have been normalized for better observation of the spectral characteristics. As shown in Fig. \ref{fig:2}, the spectrum of each harmonic from the liquids shifts progressively toward lower frequencies (i.e., a red shift) and is gradually broadened as the laser intensity increases. Whereas the corresponding harmonics from argon remain spectrally stable. Previous studies have reported frequency modulation in HHG from atomic and molecular gases, which can originate from the intrinsic changes in the medium during the strong-field interaction, such as the resonant excitation, ionization dissipation, or ultrafast nuclear motion \cite{Bian1, He1,Zuo1,Seideman1,Gibson1}, or the extrinsic factors related to the laser field itself, including its spectral chirp and pulse reshaping during propagation \cite{Kan1,Shin1}. In our experiment, the harmonic spectra from argon exhibit no discernible frequency shift across a broad range of laser intensities, thereby excluding the propagation effect in the gas HHG experiment. Furthermore, to examine potential propagation effects in the liquid HHG experiments, we have first measured the incident and transmitted spectra of the fundamental driving laser passing through the liquid flat-sheet at different intensities. The spectrum of the driving laser is demonstrated to nearly unchanged before and after the liquid sheet. In addition, we measured the harmonic spectra at different thicknesses of the liquid sheet by adjusting the vertical position of the laser focus relative to the top of the flat-sheet. The results show no significant changes in the harmonic spectra as the sheet thickness varies (see Supplementary Information). These observations rule out the influence of propagation effects on the liquid HHG spectra and indicate that the spectral modulations observed in HHG from liquid water and ethanol should originate from the intrinsic ultrafast dynamics within the liquid medium. More importantly, the spectral behavior of liquid-phase HHG differs significantly from previous results in gas medium. We observe that the redshift in liquids is consistently accompanied by substantial spectral broadening—a feature that is largely absent in gas-phase HHG. This coupled spectral modulation effect in liquid HHG also implies a distinct underlying physical mechanism of liquid HHG from gaseous HHG.

\begin{figure*}
\includegraphics[width=0.75\textwidth]{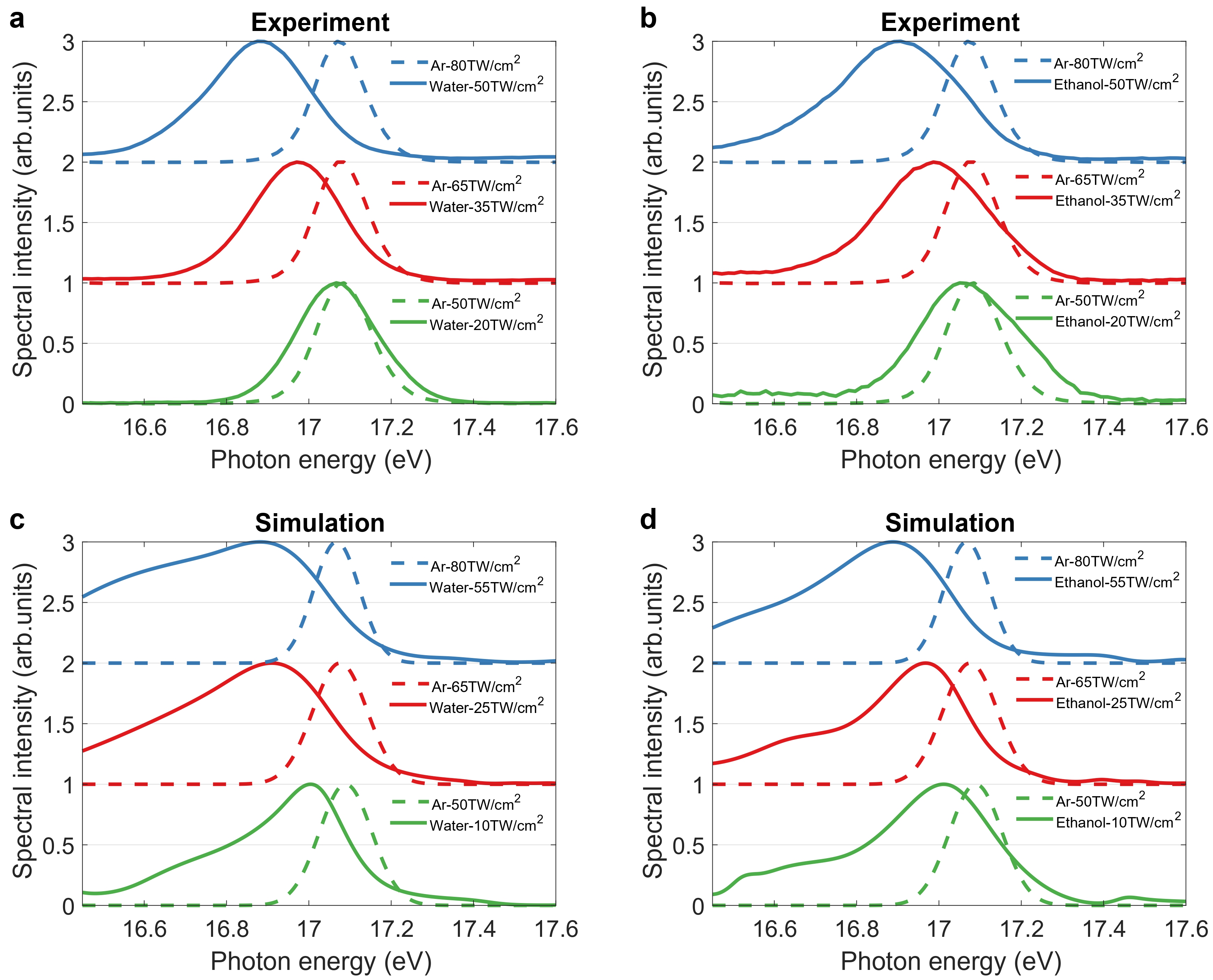}
\caption{\label{fig:3} \textbf{Measured and simulated spectra of H11 from liquids and gas.} \textbf{a}, Experimentally measured spectra (solid lines) of H11 from liquid water at some specific laser intensities. For comparison, the results of Ar gas are also presented (dashed line). \textbf{b}, Same as \textbf{a}, but for the liquid ethanol. \textbf{c-d}, Same as \textbf{a-b}, but for the simulation results. In the simulations, the laser intensities are slightly different from the experiment.}
\end{figure*}

\begin{figure}[b]
\includegraphics[width=0.4\textwidth]{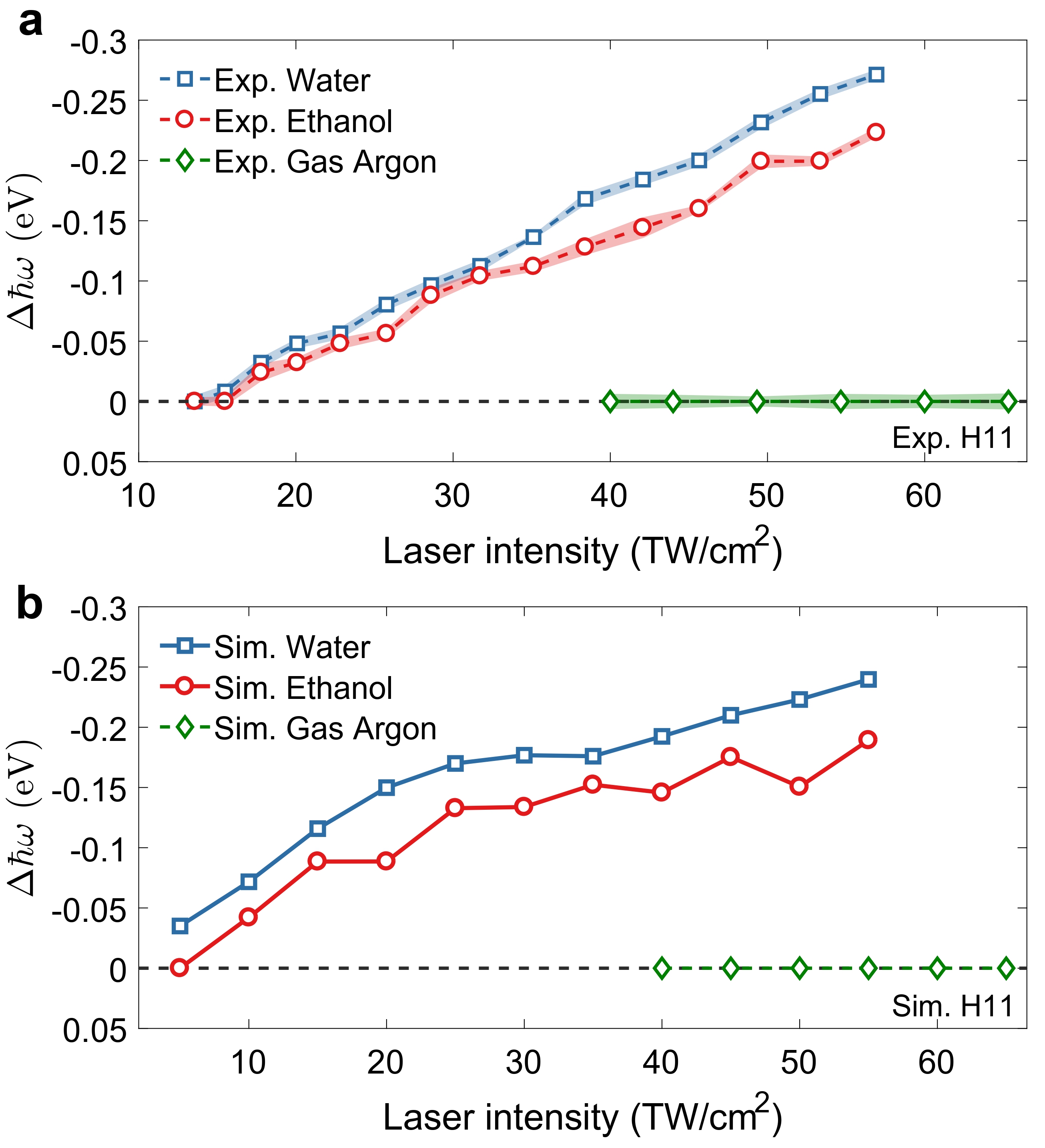}
\caption{\label{fig:4} \textbf{Laser-intensity-dependent spectral redshift of HHG from liquids and gas.} \textbf{a,} Experimentally measured spectral redshift of H11 from liquid water (squares) , ethanol (circles), and gas Ar (diamonds) as a function of the laser intensity. \textbf{b,} Same as \textbf{a}, but for the simulation results.}
\end{figure}

To understand the physical mechanism of the coupled spectral modulations in liquid HHG, we have developed an \textit{ab initio} approach for intense laser-liquid interactions. Our simulations employ periodic supercells under three-dimensional boundary conditions: (i) Deep Potential Molecular Dynamics (DPMD) for a system of 125 water molecules and (ii) \textit{ab initio} Molecular Dynamics (AIMD) for 58 ethanol molecules, both equilibrated at room temperature. More than 30 equilibrated configurations are used to reflect the macroscopic isotropy of liquids. The eigenstates of each configuration are obtained by density functional theory (DFT) calculations. The laser-liquid interaction dynamics are simulated through coupling of eigenstates with velocity-gauge time-dependent Schr\"{o}dinger equation (TDSE) solutions. Similar to the time-dependent density-functional theory (TDDFT) with frozen Kohn–Sham potentials \cite{Bauer1}, the occupied valence states below zero energy are selected as initial states. The HHG spectra are obtained by the Fourier transform of the laser-induced currents (for more details, see Methods). Although the TDSE-based independent-particle approximation neglects electron-electron interactions during time evolution, these prior studies \cite{Neufeld1,Yu1} (along with our current results) demonstrate the qualitative validity of employing simplified independent-particle models in investigating HHG in liquids. Figure \ref{fig:3} displays the experimentally measured and theoretically simulated harmonic spectra of H11 from liquid water (left panels) and ethanol (right panels) at some specific laser intensities. For comparison, the results of argon gas are also presented (dashed lines). As shown, the simulation results quantitatively reproduce the experimentally observed spectral redshift and broadening in liquid HHG. Similar results are also found for H7 and H9, which are detailed in the Supplementary Information.

For a deeper insight, we further quantitatively examined the intensity-dependent frequency redshift of HHG in liquids. Because HHG from argon gas does not have a net frequency shift, it therefore can serve as a benchmark to evaluate the frequency shift of HHG from the liquids. Figure \ref{fig:4}(a) shows the measured redshift of H11 from liquid water (squares) and ethanol (circles) as a function of the laser intensity. To quantitatively characterize the frequency shift of HHG in theory, we employed integral weights of the HHG intensity for the redshift calculation. The corresponding results are shown in Fig. \ref{fig:4}(b). One can see that the harmonic redshift in both liquids increases gradually with the laser intensity. Moreover, the redshift in water is more prominent than that in ethanol. All of these experimental findings are in reasonable agreement with the simulation results. Here, it is worth noting that accurately determining the laser intensity in strong-field experiments remains a difficult task \cite{Smeenk1,hofmann1}. In our simulations, the laser intensities used differ slightly from the experimental estimates. However, the overall agreement between the theoretical simulations and the experimental results is quite satisfactory. Our computational framework accurately resolves both the characteristic redshift and broadening observed in liquid-phase HHG spectra.


\section{Discussion}

\begin{figure}[b]
\includegraphics[width=0.5\textwidth]{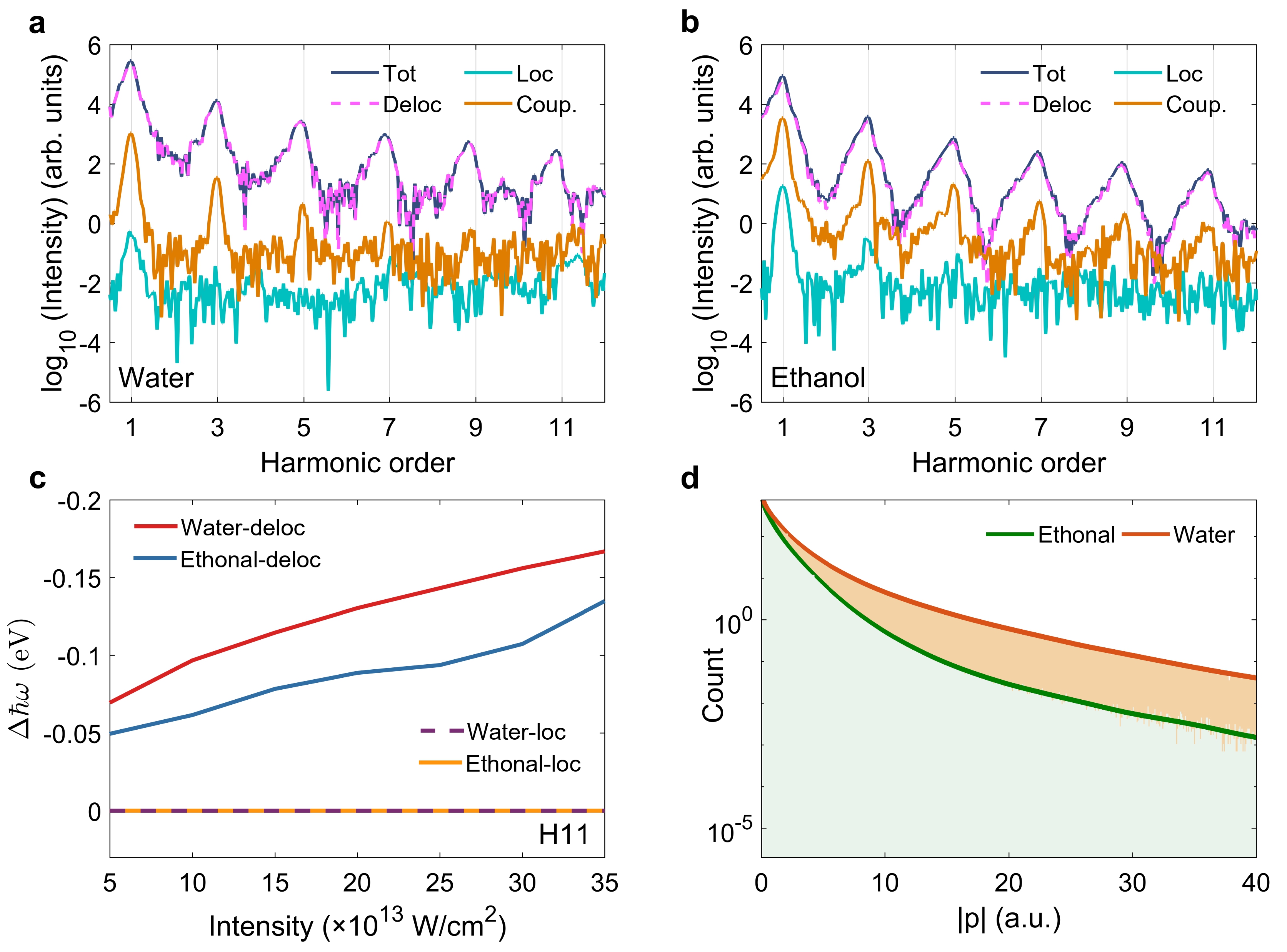}
\caption{\label{fig:5} \textbf{a, b,} Typical high-order harmonic spectra (800 nm, 55~$\text{TW}\text{/cm}^{2}$) of liquid water and ethanol obtained from simulations. \textbf{c,} Intensity-dependent frequency shift of HHG from water and ethanol obtained from the three-level transition model. \textbf{d,} Distribution of the absolute values of transition dipole moments for liquid water and ethanol.}
\end{figure}

The unique characters of harmonic spectral redshift and broadening in liquids by measurements and simulations suggest distinct physical mechanism from gaseous and solid systems. The decaying order as a function of the structure range in liquids results in the electronic states being neither fully localized nor strictly delocalized, making the analysis in real or momentum space extremely hard. To explain the underlying microscopic physics of these phenomena, we analyze the electronic dynamics in liquids in energy space. Based on the statistical distribution of the transition dipole moment $|\mathbf{p}|$, we categorize them into two types: high-$|\mathbf{p}|$ processes arising from non-local wavefunction coupling and low-$|\mathbf{p}|$ processes manifesting disorder-induced electron localization (see Methods).

Then we can write the total laser-induced current into three components: $J_D$ from the delocalized state transition with high-$|\mathbf{p}|$, $J_L$ from the localized state transition with low-$|\mathbf{p}|$, and $J_C$ from the coupling between them. Through three-component current decomposition, we systematically elucidate differentiated contributions to HHG as presented in Figs. \ref{fig:5}(a) and \ref{fig:5}(b). $J_L$ is from the highly confined localized states, and its main contribution to HHG is the lower harmonics. $J_D$ is the most predominant component for HHG. The redshift and broadening of harmonic spectra mainly come from $J_D$. The coupling $J_C$ also plays an important role, but not in the redshift and broadening.

To facilitate understanding further, we constructed an ideal three-level transition model (two bound state and one quasifree state, see Methods) to suppress complicated effects other than the transition dipole. The results can qualitatively interpret the role of large transition dipole in the redshift of HHG as illustrated in Fig. \ref{fig:5}(c). One can find that the delocalized states with high-$|\mathbf{p}|$ in this simplified model can quantitatively reproduce the laser-intensity dependent redshift, thus validating our conclusions. In contrast, no redshift occurred in $J_L$ even at significantly higher laser intensities. This finding further corroborates the crucial contribution of delocalized states to liquid harmonic generation. The more pronounced redshift in water versus ethanol in Fig. \ref{fig:4}(a) fundamentally reflects the difference between the distribution of delocalized state. As a polar liquid, water exhibits more delocalized states due to its stronger hydrogen bond network compared to weakly polar ethanol (Fig. \ref{fig:5}(d)).

An intuitive picture for the redshift is the energy loss during the transition process. Delocalized-state transition means a long path to pump the electron from one site to another. It will have more probabilities to be multiply scattered, thus leading to decrease of photon emission energy. The spectral broadening can be explained by the Stark shift which is proportional to the permanent dipole. Thus the dressed eigenvalue of the delocalized state can be easily tuned by the laser field, making the HHG spectral width larger than the localized states. We emphasize that while this simplified framework may not fully capture all physical mechanisms governing strong-field interactions in liquids, its successful reproduction of the unique redshift scaling behavior---distinct from other material phases---provides substantial heuristic value for understanding condensed matter strong-field phenomena.

\section{Conclusion}

In conclusion, our study provides direct experimental and theoretical evidence that HHG in liquids exhibits spectral behaviors fundamentally distinct from those in gases and solids, reflecting the unique electronic environment of the liquid phase. The observed redshift and broadening of harmonic spectra reveal how delocalized electrons interact with the dynamic, disordered structure of liquids, undergoing multiple scattering processes that shape their trajectories and energy loss. By incorporating multiscale \textit{ab initio} modeling, we establish a microscopic picture in which the interplay between electronic delocalization and structural fluctuations governs the nonlinear optical response. Importantly, we find that liquid polarity plays a decisive role in modulating these dynamics, offering a means of tuning the degree of electron delocalization and spectral response. These findings extend the reach of strong-field and attosecond spectroscopy into disordered media, opening a pathway for investigating ultrafast processes in complex liquid environments with high temporal and structural sensitivity. As such, this work lays the groundwork for exploring electron transport, solvation dynamics, and photochemical reactions in liquid-phase systems from a new perspective.

\section{Methods}

\subsection{Experimental set-up}

We used a commercial Ti:
sapphire laser system (Legend Elite-Duo, Coherent, Inc.), which delivers 50 fs,
800 nm pulses at a repetition rate of 1 kHz, to do the HHG experiment. A half-wave plate (HWP) in combination with a polarizer was set on the beam line to precisely adjust the laser intensity. The driving laser was focused by a $f = 40$ cm plano-convex lens onto a liquid flat-sheet in a vacuum chamber. Two liquid-nitrogen cold traps were installed to ensure a low pressure ($\textless$ 1$\times$10$^{-3}$ mbar) in the interaction chamber. The liquid flat-sheet was produced by a liquid microfluidic chip (Micro 2, Micronit, Inc.), which is pumped by a high-performance liquid chromatography (HPLC) at a flow rate of 3 mL/min.  The selection of flow rate takes into account both the thickness and stability of the liquid sheet. The microfluidic chip was mounted on a three-dimensional motorized stage
to adjust its position with respect to the laser focus.
In our experiment, the laser is focused at 3 mm below the top of the flat-sheet. The thickness of the flat-sheet is about $~1$ $\mu$m. A gas nozzle was also mounted on the same stage at a distance of 3 cm from the flat-sheet for gaseous HHG. In the gaseous HHG experiment, we adopted a low gas pressure (20 torr) and moderate laser intensities (below 100 $\text{TW}\text{/cm}^{2}$) to avoid the influence of propagation effect on the frequency shift of HHG from argon gas. The generated harmonic spectra are captured by a homemade extreme ultraviolet spectrometer \cite{NC} consisting of a slit with a width of about 0.1 mm and height of 15 mm, a flat-field grating (1200 $\text{grooves} \text{/mm}$, Shimadzu), and a microchannel plate (MCP) detector backed with a phosphor screen. The generated high-order harmonics are dispersed by the grating and imaged onto the screen. The images on the screen are read out by an
optical charge-coupled-device (CCD) camera.

\subsection{Numerical simulation}

An \textit{ab initio} computational framework for intense laser-liquid interactions is proposed to describe the HHG in liquid systems. Specifically, equilibrium configurations of periodic supercells are generated through molecular dynamics simulations. Three-dimensional supercell Deep Potential molecular dynamics (DPMD) \cite{Zhang1} simulations with 125 water molecules and \textit{ab initio} molecular dynamics (AIMD) \cite{Chen1,Li1,Lin1} simulations with 58 ethanol molecules were employed to characterize the thermodynamic equilibrium states.

The electron dynamics in the laser-liquid interaction are governed by the TDSE. Within the eigenstate basis, the TDSE in the velocity gauge is expressed as:
\begin{equation}
\begin{split}
	i\hbar\frac{\mathrm{d}}{\mathrm{d}t}c_n(t) &= \sum_m \left[ \delta_{nm}E_n + \frac{e}{m c}\mathbf{A}(t)\cdot\mathbf{p}_{nm} \right. \\
	&\quad \left. - \frac{e^2}{2m c^2}A^2(t)\delta_{nm} \right] c_m(t),
\end{split}
\end{equation}
where $c_n(t)$ denotes the expansion coefficients of the wavefunction $|\psi(t)\rangle = \sum_n c_n(t) | \phi_n \rangle$, $E_n$ is the energy of eigenstate $| \phi_n \rangle$, which is obtained in the DFT calculations of each equilibrium configuration in the above molecular dynamics simulations. Here, $\mathbf{A}(t)$ denotes the vector potential of the laser field, and $\mathbf{p}_{nm} = \langle \phi_n | \hat{\mathbf{p}} | \phi_m \rangle$ corresponds to the transition momentum matrix elements between eigenstates.

To ensure the efficiency of numerical calculations, it is necessary to truncate the infinite-dimensional Hilbert space into a finite basis set $\{\phi_n\}_{n=1}^N$. The following scheme has been devised: (i) Occupied state selection criterion: fully incorporate all occupied electron states for the expansion of the initial wave function: $|\psi(0)\rangle = \sum_{n\in {\rm occ}} c_n(0) |\phi_n\rangle$. (ii) Unoccupied state screening strategy: based on the maximum photon energy threshold $E_{\rm max} \sim 30$  eV of the target harmonic spectrum, select low-energy unoccupied states that satisfy the condition \cite{Gholam1}: $E_n \geq E_{\rm max} \geq 17\hbar\omega_L .\quad$ This criterion ensures the inclusion of transition channels up to H17.

Numerical solutions are obtained via the unconditionally stable Crank-Nicolson algorithm. The total laser-induced current $\mathbf{J}(t)$, incorporating the ensemble-averaged over $N_{\mathrm{conf}}$ configurations, is defined as:
\begin{equation}
	\mathbf{J}(t) = \frac{1}{N_{\mathrm{conf}}} \sum_{i=1}^{N_{\mathrm{conf}}} \left( -\frac{e}{m} \left\langle \psi_i(t) \left| \hat{\mathbf{p}} + \frac{e}{c}\mathbf{A}(t) \right| \psi_i(t) \right\rangle \right).
\end{equation}


The harmonic spectrum $S(\omega)$ is calculated from the Fourier transform of the current $\left| \mathcal{F}[\mathbf{J}(t)] \right|^2$. A smoothing filter is applied to $\mathbf{J}(t)$ to enhance the signal-to-noise ratio while preserving spectral peak integrity.

Calculation parameter settings as follows: (i) Laser intensity $I_0 = 10\text{-}60~\text{TW}~\text{cm}^{-2}$, central wavelength $\lambda=800$ nm ($\hbar\omega=1.55$ eV), total pulse duration 80 fs ($T\approx30$ cycles), time step $\Delta t=0.1$ a.u., with a sine-squared envelope shape. (ii) 1600 normalized eigenstates (with 500-600 initially occupied states \cite{Zhao1}).

Here, we present a concise explanation of how to distinguish between the two types of transition dipole moments \( |\mathbf{p}| \). Based on the spatial distribution of the electronic wave function in the liquid, regions with \( |\mathbf{p}| > 6 \) a.u. correspond to the high-\( |\mathbf{p}| \) type from the transition of delocalized states, and regions with \( |\mathbf{p}| < 6 \)  a.u. correspond to the low-\( |\mathbf{p}| \) type from the transition of localized states. Therefore, we can clearly classify them into two categories.

To further filter the physics behind the liquid HHG, we also employ a three-level transition model for qualitative analysis of harmonic redshift. The framework is elaborated as follows: (i) Energy-level pairing and independent evolution: two bound states $E_j$, $E_k$, and one quasi-free state $E_l$. Solve the TDSE for each $\mathcal{S}_{jkl}$ subsystem. (ii) Coherent superposition: calculate microscopic current contributions from individual three-level systems, then obtain high-harmonic spectra through Fourier transformation after coherently superimposing currents from all subsystems.

\begin{acknowledgments}
We acknowledge the Public Service Platform of High Performance Computing provided by Network and Computing Center of Huazhong University of Science and Technology (HUST) for providing the computing resources.
\end{acknowledgments}

\section*{Author Contributions} 
 W.T. and Z.D. contributed equally to this letter. L.H., P.L. and P.L. conceived the project. W.T. and L.H. designed and carried out the experiment. Z.D. and X.B. developed the theory and performed the numerical simulations. All authors participated in the discussion of the results and contributed to the manuscript. 

\section*{Funding}
We gratefully acknowledge funding from National Key Research
and Program of China
(Grant No. 2023YFA1406800), and National Natural Science
Foundation of China (Grants No. 12225406, No. 12274421, No. 12074136, No. 11934006, and No. 12021004).

\section*{Conflicts of Interest}
The authors declare that there is no conflict of interest regarding the publication of this article.

\section*{Data Availability}
The data that support the findings of this study are available from the corresponding
authors upon reasonable request.

\nocite{*}


\end{document}